\documentclass{emulateapj}

\newcommand{\etal}{et al.~}

\newcommand{\hbetao}{H$\beta_{\rm o}$}
\newcommand{\hbetal}{H$\beta_{\rm LICK}$}
\newcommand{\hbeta}{H$\beta$}
\newcommand{\hgamma}{H$\gamma$}
\newcommand{\hdelta}{H$\delta$}

\shorttitle {Blue stragglers rejuvenating globular clusters} \shortauthors {A.~J.~Cenarro et al.}

\begin{document}

\title {Evidence for blue straggler stars rejuvenating the integrated
  spectra of globular clusters}

\author{A. Javier Cenarro, J.L. Cervantes, Michael A. Beasley, Antonio Mar\'in--Franch, Alexandre Vazdekis}
\affil{Instituto de Astrof\'isica de Canarias,  V\'ia L\'actea s/n, 38200 La Laguna, 
Tenerife, Spain}
\email{cenarro@iac.es}

\begin{abstract}
Integrated spectroscopy is the method of choice for deriving the ages
of unresolved stellar systems. However, hot stellar evolutionary
stages, such as hot horizontal branch stars and blue straggler stars
(BSSs), can affect the integrated ages measured using Balmer
lines. Such hot, ``non-canonical'' stars may lead to overestimations
of the temperature of the main sequence turn-off, and therefore
underestimations of the integrated age of a stellar population. Using
an optimized \hbeta\ index in conjunction with HST/WFPC2
color-magnitude diagrams (CMDs), we show that Galactic globular
clusters exhibit a large scatter in their apparent ``spectroscopic''
ages, which does not correspond to that in their CMD-derived ages. We
find for the first time that the specific frequency of BSSs, defined
within the same aperture as the integrated spectra, shows a clear
correspondence with \hbeta\ in the sense that, at fixed metallicity,
higher BSS ratios lead to younger $apparent$ spectroscopic ages. Thus,
the specific frequency of BSSs in globular clusters sets a fundamental
limit on the accuracy for which spectroscopic ages can be determined
for globular clusters, and maybe for other stellar systems like
galaxies. The observational implications of this result are discussed.
\end{abstract}


\section{Introduction} \label{Introduction}

A common method for estimating the ages of unresolved stellar systems
is to measure Balmer lines and metal lines from integrated spectra,
and compare them to stellar population models. The method relies on
the fact that Balmer lines are mostly sensitive to the effective
temperature ($T_{\rm eff}$) of the main sequence (MS) turn-off of a
stellar population (e.g.~Buzzoni \etal1994). This, in turn, provides a
measure of the population age. 

However, a longstanding concern over the use of Balmer lines to
estimate spectroscopic ages is the effect of non-canonical
evolutionary stages on the integrated stellar population spectra.
Particularly worrisome are the possible effects that bluer horizontal
branch (HB) stars and blue straggler stars (BSSs) may have on the
inferred $T_{\rm eff}$ of the MS turn-off. HB stars are helium-burning
stars which occupy a region in the CMD with a typical absolute $V$
magnitude of $\sim 0.7$\,mag (e.g.~Harris 2001). BSSs are identified
as blue, luminous extensions of MS stars (Sandage 1953). They are
thought to form due to H refuelling processes after the MS stage,
either from collisional stellar encounters (e.g.~Davies \etal1994) or
from mass-transfer binaries (McCrea 1964). Both HB stars and BSSs with
$T_{\rm eff} > 6500$\,K show prominent Balmer lines in their spectra,
which can mimic the presence of younger stellar populations (e.g.~Rose
1985, Lee \etal2000, Schiavon \etal2004; Trager \etal2005).

In a recent paper, Cervantes \& Vazdekis (2008) defined an optimized
line-strength index for \hbeta, called \hbetao, that minimizes the
metallicity dependence of the Balmer line in favor of its age
sensitivity. Interestingly, as pointed out in that paper, the
integrated spectra of Galactic globular clusters (GGCs) from Schiavon
\etal(2005; hereafter S05) exhibit a clear intrinsic scatter in their
\hbetao\ strengths, particularly with increasing metallicity.

In order to shed light on the above issue, in this {\it letter} we
combine resolved and unresolved data of such GGCs to investigate
whether CMD-based age differences among GGCs and/or different relative
contributions of non-canonical stages can be responsible for the
distinct integrated \hbetao\ indices. In Section~\ref{TheData} we
discuss the data used in this study, the analysis of which is
described in Section~\ref{Analysis}. Finally, in
Section~\ref{ConclusionsandDiscussion} we present our conclusions and
discuss our findings within the context of estimating ages for
unresolved stellar populations.

\section{The Data} \label{TheData}

Integrated optical spectra for 41 GGCs were taken from S05. These data
were obtained by drift-scanning the core diameter of each GGC --with a
spectroscopic aperture equal to this diameter-- in order to construct
a representative integrated spectrum. The spectra cover a wavelength
range of $\sim3350-6430$\AA\ with a FWHM of $\sim 3.1$\AA\ and typical
signal-to-noise ratios (S/N) of $\sim 250$\,\AA$^{-1}$ in the
continuum of the \hbeta\ line. We refer the reader to S05 for more
details on these data.

In Figure~\ref{fig1} we present the indices \hbetao, \hbetal\ (Worthey
\etal1994) and [MgFe] (Gonz\'alez 1993) measured for the 41 GGC
spectra at the S05 spectral resolution. Uncertainties in the index
measurements (1$\sigma$ error bars) account for the S/N spectra and
the typical radial velocity error provided by S05 for each GGC. To
guide the eye, based on the MILES stellar library
(S\'anchez-Bl\'azquez \etal2006a; Cenarro \etal2007a), an extension of
the simple stellar population (SSP) models in Vazdekis (1999) are
overplotted at 3.1\AA\ spectral resolution (hereafter V99$+$). The
fact that most GGCs lie below the model grids arises from the
well-known zero-point problem of SSP models (e.g.~Vazdekis \etal2001;
Schiavon \etal2002), although this does not affect our results which
are based on relative differences.

As expected, in the \hbetao\ plot (Fig.~\ref{fig1}a) GGCs follow --on
average-- an old sequence in metallicity. However, by simple visual
inspection they seem to separate into two groups, particularly at the
high metallicity end ([MgFe] $\gtrsim 1.2$). Hence, solid and open
symbols are employed to indicate, respectively, GGCs with $high$ and
$low$ \hbetao\ values at fixed [MgFe]. Henceforth, these are referred
to as GGC$H$ and GGC$L$ respectively.

When using the \hbetal\ definition (Fig.~\ref{fig1}b), both GGC groups
are still distinguished, although the greater age--metallicity
degeneracy of \hbetal\ would make their differentiation less clear to
detect if the distinct symbol codes were not present. Even so, Puzia
\etal(2002) already pointed out the existence of unexpected \hbetal\
differences among certain metal rich (MR) GGCs. It therefore appears
that some property differs between both groups leading to different
Balmer-line strengths at a given metallicity. In fact, since the
effect seems to increase with the increasing metallicity, we focus
our analysis on those 25 GGCs from S05 with [Fe/H] $> -1.35$ ([MgFe]
$\gtrsim 1.2$) among which clear differences in \hbeta\ at fixed
metallicity are observed.

Together with the indices in Fig.~\ref{fig1}, the adopted CMD-derived
parameters for the 25 GGCs are listed in Table~\ref{table}. The HB
morphology, as measured by the HBR parameter (Lee \etal1994), and the
specific frequency of RR Lyrae variables, $S_{\rm RR Lyrae}$, are
taken from the Harris (1996) catalogue (Feb 2003 revision; hereafter
H03). We adopted the relative age estimates from De Angeli \etal(2005)
and Recio-Blanco \etal(2006), whose applied the so-called vertical
method on the GGC HST/WFPC2 snapshot catalogue of Piotto
\etal(2002). Also based on that catalogue, Recio-Blanco \etal(2006)
estimated the maximum $T_{\rm eff}$ along the HB, $T_{\rm eff}\,{\rm
HB}$ --considered as an HB morphology parameter--, whereas Moretti
\etal(2008; hereafter M08) measured the logarithm of the number of
BSSs inside the core radius, $r_{\rm c}$, normalised to the sampled
luminosity --in units of $10^4$\,L$\sun$-- in the $F555W$ HST band
within the same aperture. The last quantity can be considered as a
logarithmic specific frequency of BSSs inside the GGC $r_{\rm c}$,
hereafter $S_{\rm BSS}^{r_{\rm c}}$, and is representative of the
spectroscopic data in S05 as they are both computed within the same
aperture.

\section{Analysis} 
\label{Analysis}

With the aim of constraining the origin of the intrinsic scatter in
the Balmer line-strengths of our GGC subsample, in Figure~\ref{fig2}
we show the CMD-derived parameters of Section~\ref{TheData}
--where available-- as a function of the GGC metallicity from H03. 
Symbol codes are kept as in Fig~\ref{fig1} except for the MR GGC$H$s
NGC6388 and NGC6441, which being well-known {\it second parameter}
clusters (Rich \etal1997), are plotted as open stars --rather than
solid squares-- to facilitate further discussion.

In Fig.~\ref{fig2}a, we see that there is no dependence of the group
location on the CMD-derived age. The difference in \hbeta\ strengths
between both groups is therefore not due to age differences among the
GGCs, as would normally be inferred from a classical SSP index-index
analysis. Note also that the typical dispersion of $\sim 1$\,Gyr
quoted by De Angeli \etal(2005) among the CMD-derived ages of
intermediate metallicity GGCs could never explain the large scatter in
\hbeta. We also rule out the possibility that the number of RR lyraes
in the instability strip is playing a significant role, as the two GGC
groups are well mixed (Fig.~\ref{fig2}b). In addition we find that
there is no obvious dependence on the HB morphology, as measured by
either the HBR parameter or the maximum $T_{\rm eff}$ of the HB
(Fig.~\ref{fig2}c and Fig.~\ref{fig2}d, respectively). As expected,
the second parameter clusters NGC6388 and NGC6441 (open stars) do
stand out of the general trends in both panels. Their high \hbeta\
strengths are naturally explained by the addition of hot HB stars in
their integrated spectra.

Besides basic properties of the GGC CMDs, literature estimates of
[$\alpha$/Fe] for the GGCs in S05 (where available) show a high degree
of homogeneity (e.g.~Pritzl \etal2005), so differing levels of
$\alpha$--elements cannot account for the observed differences.  We
have also ruled out that the Balmer lines of GGC$L$s are
sistematically filled-in by emission. In fact, emission was found and
corrected by S05 for only NGC6171 and NGC6553 (GGC$H$s), and NGC6352
(GGC$L$). Interestingly, the fact that NGC6352 has the weakest \hbeta\
line of the sample may suggest that a residual emission could still be
present.

Having rejected the above mechanisms from being responsible for the
observed differences in \hbeta\ between GGC$H$s and GGC$L$s --except
in the obvious case of NGC6388 and NGC6441--, in Fig~\ref{fig2}e we
show the GGC metallicities versus $S_{\rm BSS}^{r_{\rm
c}}$. Interestingly, GGC$H$s and GGC$L$s --which were identified
spectroscopically-- separate cleanly into two groups in terms of their
BSS specific frequencies. At a given metallicity, GGCs with higher
$S_{\rm BSS}^{r_{\rm c}}$ values exhibit stronger \hbeta\ lines,
suggesting that BSSs are indeed affecting their integrated spectra.

To reinforce this result, we quantify the impact of BSSs on the
integrated spectrum of NGC6342, the GGC$H$ with the highest $S_{\rm
BSS}^{r_{\rm c}}$. Based on photometric data from H03 and M08 for this
GGC and its BSS population, we obtain that $13\%$ of the GGC flux in
$V$ band within $r_{\rm c}$ comes from a population of 7 BSSs with
$0.22 \leq B-V \leq 0.52$, and a luminosity-weighted $B-V$ of $0.33$.
For each BSS, assuming [Fe/H] $= -0.65$ and its $B-V$, we estimate
$T_{\rm eff}$ using the $(B-V)-T_{\rm eff}$ relation for dwarfs from
Alonso \etal(1996). The $T_{\rm eff}$ values of the 7 BSSs are in the
range $5925 \leq T_{\rm eff} \leq 7500$\,K. We then compute spectral
BSS templates with the above parameters on the basis of the MILES
stellar library, using the interpolating algorithm in Vazdekis
\etal(2003; Appendix B). After the 7 BSS templates are scaled
according to their $V$ fluxes and subtracted from the NGC6342
spectrum, its integrated \hbetao\ and \hbetal\ indices decrease by
$0.70$\,\AA\ and $0.62$\,\AA\ respectively. Taking into account that
the averaged $S_{\rm BSS}^{r_{\rm c}}$ of GGC$L$s with [Fe/H]$ > -0.8$
(similar to that of NGC6342) is $\sim 1.4$ (Fig.~\ref{fig2}e), and
assuming similar $T_{\rm eff}$ values for their BSS populations, the
relative offsets in \hbetao\ and \hbetal\ between NGC6342 and the
above GGC$L$s are, respectively, $0.58$\,\AA\ and $0.51$\,\AA. This
agrees with the differences between GGC$H$s and GGC$L$s obtained at
this metallicity regime ([MgFe]$\gtrsim 2$; Fig.~\ref{fig1}),
supporting the idea that BSSs are responsible for the observed
differences in \hbeta. Even more, the effect of BSSs is also detected
among the subsample of GGC$L$s with [Fe/H]$ > -0.8$. NGC0104 and
NGC6624, with extreme values of $S_{\rm BSS}^{r_{\rm c}}$, pose the
lowest and largest \hbetao\ values respectively.

\section{Discussion} 
\label{ConclusionsandDiscussion}

Based on the close correspondence between the specific frequency of
BSSs in GGCs with [Fe/H] $> -1.35$ and their integrated \hbeta\
strengths at fixed metallicity, we conclude that BSSs are primarily
responsible for the \hbeta\ variations observed in the integrated
spectra of GGCs of intermediate-to-high metallicity. Far from
discussing on the origin for the distinct $S_{\rm BSS}^{r_{\rm c}}$
values among GGCs (see M08 for a thorough study on this topic), we
here analyze the implications of the above result in the context of
age-dating unresolved stellar populations.

\begin{figure}
%\epsscale{0.65}
\epsscale{0.45}
\plotone{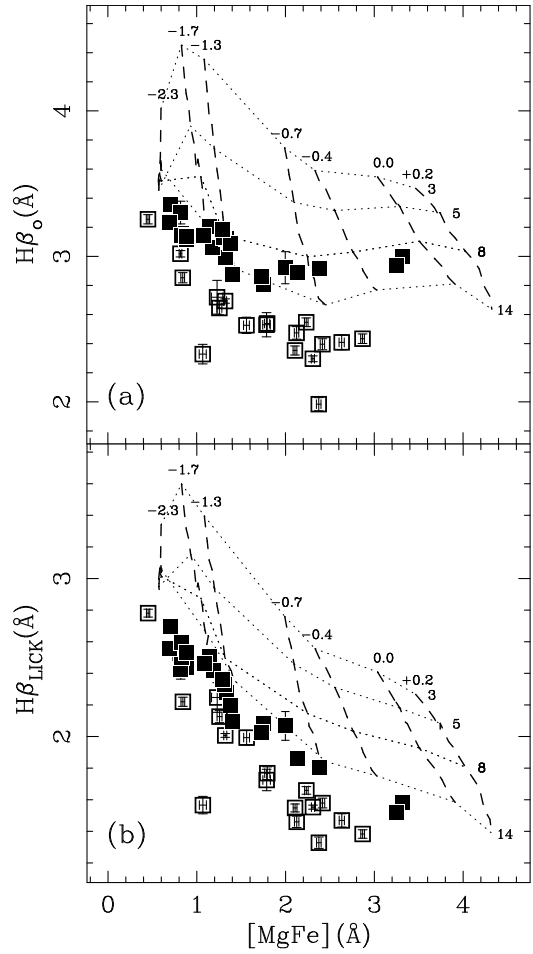}
\caption{\hbetao\ (a) and \hbetal\ (b) indices versus [MgFe] for the
  41 GGCs in S05. Solid and open symbols indicate, respectively, GGCs
  with high (GGC$H$) and low (GGC$L$) \hbeta\ indices at fixed
  [MgFe]. SSP models from V99$+$ at FWHM = 3.1\AA\ are overplotted,
  with dotted and dashed lines indicating fixed ages and
  metallicities as in the labels.}
\label{fig1}
\end{figure}

First, caution must be employed in Balmer-line based age-metallicity
studies of unresolved extragalactic globular clusters (EGCs). Cenarro
\etal(2007b) already reported the existence of EGCs with strong Balmer
lines that were consistent with hosting an additional population of
either blue HB stars and/or BSSs. Since the BSS fraction of EGCs is
generally not known, the finding in this letter sets a fundamental
limit to the reliability with which ages may be determined for EGCs
using Balmer lines and SSP models. Taking the S05 data as a
representative old GC system, we can estimate this limit from the
averaged offsets in the measured \hbeta\ lines of GGC$H$s and
GGC$L$s. Since the offsets seem to vary with metallicity, local linear
fits to all GGC$H$s and GGC$L$s in Fig.~\ref{fig1} with [MgFe]$> 1.5$
([Fe/H] $\gtrsim -1.0$) have been performed. For instance, at the
location of 47Tuc (NGC0104; [MgFe]$\sim 2.31$), we obtain
$\Delta$\hbetao\ $= 0.46 \pm 0.03$\AA\ and $\Delta$\hbetal\ $= 0.33
\pm 0.04$\AA\ (at the S05 resolution) with uncertainties accounting
for the standard errors of the means. Thus, assuming that GGC$L$s are
$\sim14$\,Gyr old --the largest SSP age in Fig.~\ref{fig1}-- the two
offsets can be consistently misinterpreted on the basis of SSP models
as GGC$H$s being $\sim 6-7$\,Gyr old, that is, as a rejuvenation of up
to $\sim 8$\,Gyr. Differences between GGC$H$s and GGC$L$s also exist
for the Lick \hgamma\ and \hdelta\ indices (Worthey \& Ottaviani
1997), although they are not so apparent, probably due to their
limited age-disentangling power for old SSPs. For these indices, the
above test leads to rejuvenations of up to $\sim 4 - 5$\,Gyr.

\begin{figure*}
\epsscale{1.19}
%\epsscale{1.}
\plotone{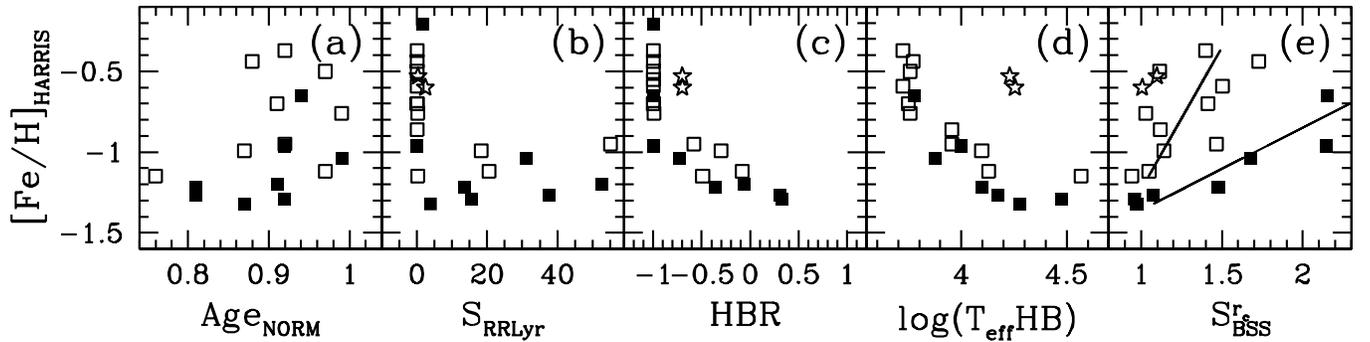}
\caption{Metallicity versus CMD-derived parameters (Age$_{\rm NORM}$,
  $S_{\rm RRLyr}$, HBR, $log(T_{\rm eff}\,{\rm HB})$, and $S_{\rm
  BSS}^{r_{\rm c}}$; see definition in the text) for a subsample of 25
  GGCs from S05 with [Fe/H]$ > -1.35$. Symbol codes are kept as in
  Fig~\ref{fig1} except for the second parameter GGCs NGC6388 and
  NGC6441 (open stars). Solid lines in panel $e$ illustrate individual
  linear fits to solid and open squares.}
\label{fig2}
\end{figure*}

The role of metallicity in the present discussion is worthwhile
considering. Although the relation between $S_{\rm BSS}^{r_{\rm c}}$
and metallicity is not statistically significant --but marginally
positive-- over the entire GGC sample (in agreement with M08), we find
clear correlations for GGC$H$s and GGC$L$s separately, as illustrated
by the solid lines in Fig.~\ref{fig2}e. The different slopes seem to
indicate that, when BSSs are important, their relative contributions
are larger at high metallicities. Note that the fading with
metallicity expected in $F555W$ band for the most MR GGCs only
accounts for up to $\Delta S_{\rm BSS}^{r_{\rm c}} \sim 0.2$, so the
above trends are irrespective of this effect.

These results may also have important consequences for EGC studies. To
understand the origin of color bimodality in GC systems within a
context of galaxy formation, age-dating GC subpopulations --through
the analysis of their integrated Balmer lines-- is a common practice
(see Brodie \& Strader 2006 and references therein). Interestingly,
some papers have reported that the MR GC subpopulation of certain
galaxies show on average a smaller mean age and a larger age scatter
than their metal poor (MP) counterparts (e.g.~Puzia
\etal2005). Although the present finding does not rule out the
existence of {\it true} age differences between MP and MR GC
subpopulations, the increasing importance of BSSs with metallicity
might, at least, partially affect the results of previous work.

Whether all the above results can compromise the integrated ages of
other stellar systems, like galaxies, may rely on the mechanism that
dominates the formation of BSSs. If stellar encounters were driving
the BSS population, then one should not expect a major effect in
galaxies because of their much lower stellar densities. However, this
would not apply if mass-transfer binaries were the progenitors of most
BSSs. In fact, Momany \etal(2007) and Mapelli \etal(2007) support the
last scenario to explain the large BSS populations of dwarf spheroidal
galaxies, and Han \etal(2007) have demonstrated the importance of
binary interactions to understand the UV-upturn of elliptical galaxies
(Es). It therefore seems that BSSs could play a non-negligible role
in the integrated spectra of galaxies as long as they host an
important fraction of binary stars. If this were the case and the
potential increasing importance of BSSs with metallicity would still
hold for massive Es, then BSSs could contribute to the age scatter
reported for massive Es and to the fact that $younger$ Es have higher
metallicities than $older$ Es (e.g.~Trager \etal2000;
S\'anchez-Bl\'azquez \etal 2006b). This picture, however, requires of
further investigation which is out of the scope of this paper.

\acknowledgments We acknowledge the anonymous referee for very useful
comments. A.J.C. and A.M-F are Juan de la Cierva Fellows of the
Spanish Ministry of Education and Science (SMES). JLC acknowledges the
SMES for a FPU PhD fellowship. This work has been funded by the SMES
through grant AYA2007-67752-C03-01.

{}

%\clearpage

\begin{deluxetable}{ccccccccccc}
\tabletypesize{\scriptsize}
\tablecaption{Galactic globular clusters in Schiavon \etal(2005) with [Fe/H] $> -1.35$ \label{table1}}
\tablewidth{0pt}
\tablehead{
  \colhead{GGC} &  
  \colhead{Group\tablenotemark{a}} & 
  \colhead{\hbetao\tablenotemark{b}} &  
  \colhead{\hbetal\tablenotemark{b}}  &  
  \colhead{[MgFe]\tablenotemark{b}} &
  \colhead{[Fe/H]\tablenotemark{c}} &  
  \colhead{Age$_{\rm NORM}$\tablenotemark{d}} &
  \colhead{$S_{\rm RRLyr}$\tablenotemark{c}} & 
  \colhead{HBR\tablenotemark{c}} &  
  \colhead{$log(T_{\rm eff}\,{\rm HB})$\tablenotemark{e}} &  
  \colhead{$S_{\rm BSS}^{r_{\rm c}}$\tablenotemark{f}}
 }
\startdata
NGC0104 & L & $2.30\pm0.02$ & $1.55\pm0.01$ & $2.31\pm0.02$ & $-0.76$ & 0.99 &  0.2 & $-0.99$                 &  3.756 & 1.03 \\
NGC1851 & H & $2.99\pm0.03$ & $2.27\pm0.02$ & $1.32\pm0.02$ & $-1.22$ & 0.81 & 13.5 & $-0.36$                 &  4.097 & 1.48 \\
NGC2808 & L & $2.69\pm0.01$ & $2.01\pm0.01$ & $1.32\pm0.02$ & $-1.15$ & 0.76 &  0.3 & $-0.49$                 &  4.568 & 0.94 \\
NGC5904 & H & $3.06\pm0.01$ & $2.42\pm0.01$ & $1.18\pm0.01$ & $-1.27$ & 0.81 & 37.7 & $+0.31$                 &  4.176 & 1.07 \\
NGC5927 & L & $2.43\pm0.03$ & $1.38\pm0.03$ & $2.87\pm0.02$ & $-0.37$ & 0.92 &  0.0 & $-1.00$                 &  3.724 & 1.40 \\
NGC6121 & H & $3.12\pm0.03$ & $2.28\pm0.03$ & $1.34\pm0.02$ & $-1.20$ & 0.91 & 52.7 & $-0.06$                 &   --   &  --  \\
NGC6171 & H & $2.81\pm0.05$ & $2.08\pm0.05$ & $1.76\pm0.03$ & $-1.04$ & 0.99 & 31.0 & $-0.73$                 &  3.875 & 1.68 \\
NGC6266 & H & $2.88\pm0.02$ & $2.09\pm0.02$ & $1.41\pm0.02$ & $-1.29$ & 0.92 & 15.6 & $+0.32$                 &  4.477 & 0.96 \\
NGC6284 & H & $3.13\pm0.04$ & $2.33\pm0.04$ & $1.31\pm0.03$ & $-1.32$ & 0.87 &  3.9 &  --                     &  4.279 & 0.97 \\
NGC6304 & L & $2.41\pm0.05$ & $1.47\pm0.04$ & $2.63\pm0.03$ & $-0.59$ &  --  &  0.0 & $-1.00$                 &  3.724 & 1.51 \\
NGC6316 & L & $2.47\pm0.04$ & $1.46\pm0.04$ & $2.13\pm0.02$ & $-0.55$ &  --  &  --  & $-1.00$                 &   --   &  --  \\
NGC6342 & H & $2.92\pm0.11$ & $2.07\pm0.09$ & $2.00\pm0.06$ & $-0.65$ & 0.94 &  --  & $-1.00$                 &  3.778 & 2.16 \\
NGC6352 & L & $1.98\pm0.05$ & $1.33\pm0.04$ & $2.37\pm0.03$ & $-0.70$ &  --  &  0.0 & $-1.00$                 &   --   &  --  \\
NGC6356 & L & $2.40\pm0.04$ & $1.58\pm0.03$ & $2.42\pm0.02$ & $-0.50$ & 0.97 &  0.0 & $-1.00$                 &  3.756 & 1.12 \\
NGC6362 & L & $2.53\pm0.06$ & $1.99\pm0.05$ & $1.56\pm0.03$ & $-0.95$ & 0.92 & 55.1 & $-0.58$                 &  3.954 & 1.47 \\
NGC6388 & H & $2.89\pm0.02$ & $1.86\pm0.02$ & $2.14\pm0.02$ & $-0.60$ &  --  &  2.4 & $-0.70$\tablenotemark{g}&  4.255 & 1.01 \\
NGC6441 & H & $2.91\pm0.03$ & $1.80\pm0.02$ & $2.38\pm0.02$ & $-0.53$ &  --  &  0.3 & $-0.70$\tablenotemark{g}&  4.230 & 1.10 \\
NGC6528 & H & $3.00\pm0.04$ & $1.59\pm0.03$ & $3.32\pm0.02$ & $-0.04$ &  --  &  0.0 & $-1.00$                 &   --   &  --  \\
NGC6553 & H & $2.94\pm0.05$ & $1.52\pm0.04$ & $3.25\pm0.03$ & $-0.21$ &  --  &  1.6 & $-1.00$                 &   --   &  --  \\
NGC6569 & L & $2.53\pm0.08$ & $1.73\pm0.07$ & $1.79\pm0.04$ & $-0.86$ &  --  &  0.0 &  --                     &  3.954 & 1.12 \\
NGC6624 & L & $2.55\pm0.03$ & $1.66\pm0.02$ & $2.23\pm0.02$ & $-0.44$ & 0.88 &  0.0 & $-1.00$                 &  3.771 & 1.73 \\
NGC6637 & L & $2.35\pm0.03$ & $1.55\pm0.03$ & $2.11\pm0.02$ & $-0.70$ & 0.91 &  0.0 & $-1.00$                 &  3.748 & 1.41 \\
NGC6638 & L & $2.54\pm0.05$ & $1.77\pm0.04$ & $1.79\pm0.03$ & $-0.99$ & 0.87 & 18.3 & $-0.30$                 &  4.097 & 1.14 \\
NGC6652 & H & $2.86\pm0.03$ & $2.03\pm0.02$ & $1.73\pm0.02$ & $-0.96$ & 0.92 &  0.0 & $-1.00$                 &  4.000 & 2.14 \\
NGC6723 & L & $2.64\pm0.05$ & $2.13\pm0.04$ & $1.26\pm0.03$ & $-1.12$ & 0.97 & 20.5 & $-0.08$                 &  4.130 & 1.05 \\
\enddata
\tablenotetext{a}{GGCs with high (H) and low (L) \hbeta\ indices at fixed metallicity.} 
\tablenotetext{b}{Measured at FWHM = 3.1\AA\ spectral resolution.}
\tablenotetext{c}{From the Harris (1996, February 2003 revision) catalogue.}
\tablenotetext{d}{Relative GGC ages from De Angeli \etal(2005) and Recio-Blanco \etal(2006).}
\tablenotetext{e}{Maximum $T_{\rm eff}$ along the HB, from Recio-Blanco \etal(2006).}
\tablenotetext{f}{$S_{\rm BSS}^{r_{\rm c}} = log(N_{\rm BSS}/L_{\rm F555w})$ inside the core radius. $L_{\rm F555w}$ in units
of $10^4$\,L$\sun$. Taken from Moretti \etal(2008).}
\tablenotetext{g}{NGC6388 value taken from Zoccali \etal(2000). The same value is assumed for NGC6441 (Puzia \etal2002).}
\label{table}
\end{deluxetable}

\end{document}